\documentclass[english,a4paper,12pt]{article}
\usepackage[T2A]{fontenc}
\usepackage[cp1251]{inputenc}
\usepackage{babel}
\usepackage{amsmath}
\usepackage{graphicx}
\usepackage{amssymb}
\usepackage{epsf}
\usepackage{pazhe}
\usepackage[labelsep=period,hang]{caption} 

\tightenlines

\newcommand{\WI}[2]{#1_{\mathrm{#2}}}
\newcommand{\rhob}{\WI{\rho}{b}}
\newcommand{\rhod}{\WI{\hat{\rho}}{b}}
\newcommand{\mb}{\WI{m}{b}}
\newcommand{\rg}{\WI{r}{g}}
\sloppypar

\begin{document}
	\baselineskip 21pt
	
	
	\title{\bf Explosion of a Minimum-Mass Neutron Star within Relativistic Hydrodynamics}
	
	\author{\bf \hspace{-1.3cm}\copyright\, 2022 г. \ \
		A.V. Yudin\affilmark{1}}
	
	\affil{
		{\it National Research Center ``Kurchatov Institute'', Moscow, 123182 Russia}$^1$}
	
	\vspace{2mm}
	
	\sloppypar
	\vspace{2mm}
	\noindent
	The relativistic hydrodynamics equations are adapted for the spherically symmetric case and
the Lagrangian form. They are used to model the explosive disruption of a minimum-mass neutron star, a
key ingredient of the stripping model for short gamma-ray bursts. The shock breakout from the neutron star
surface accompanied by the acceleration of matter to ultrarelativistic velocities is studied. A comparison
with the results of previously published nonrelativistic calculations is made.

	\noindent
	{\bf Keywords:\/} neutron stars, relativistic hydrodynamics, shock waves, gamma-ray bursts.
	
	\noindent
	
	\vfill
	\noindent\rule{8cm}{1pt}\\
	{$^*$ email: $<$yudin@itep.ru$>$}
	
	\clearpage
	
\section*{INTRODUCTION}
\noindent On August 17, 2017, the event GW170817 with
parameters corresponding to merging neutron stars
occurred on the LIGO and Virgo gravitational-wave
antennas (Abbott et al. 2017). In addition, the
FERMI and INTEGRAL satellites detected the accompanying
gamma-ray burst GRB170817A almost
simultaneously. Thus, the connection between the
merging of neutron stars (NSs) and short gammaray
bursts (GRBs) (Blinnikov et al. 1984) was directly
confirmed for the first time.

However, this GRB turned out to be peculiar,
which renewed the nearly faded interest in the stripping
model for short GRBs (Blinnikov et al. 2021).
In contrast to the universally accepted NS merging
model, in the stripping model two NSs, having
come closer together due to the energy losses through
gravitational-wave radiation, do not merge, but begin
to exchange mass, with the more massive one
swallows (strips) its less massive companion (Clark
and Eardley 1977). The latter, having reached a
configuration corresponding to the minimum mass of
stable NSs ($\sim 0.1M_\odot$), explodes to produce a GRB.

The details of the explosion of a minimum-mass
NS (MMNS) were first calculated by Blinnikov
et al. (1990) (see also Sumiyoshi et al. 1998). An
important fact for us here is that the matter expansion
velocities after the explosion are, on average, $\sim$10\%
of the speed of light. In addition, the shock from
the explosion, when breaking out, is accelerated as
it propagates along a descending density profile. A
computation of this process within nonrelativistic
hydrodynamics with a good resolution can lead and
actually leads, as we will show, to velocities exceeding
the speed of light. Therefore, studying this process in
terms of relativistic hydrodynamics is topical. In this
case, the gravitational field may be deemed relatively
weak: the MMNS has a factor of 10 lower mass and
a factor of 10 larger radius than does an ordinary
NS; consequently, the general relativity (GR) effects
for it are a factor of 100 smaller than those for an
ordinary NS. It seems to us that the most appropriate
approach to study this problem is the approximation
proposed in Hwang and Noh (2016), which we will
use.
	
\section*{BASIC EQUATIONS}
\noindent Let us write the relativistic hydrodynamics equations
following Hwang and Noh (2016). The gravitational
field is assumed to be weak (i.e., the GR
effects are small), but the matter velocities and the
energy density in comparison with the rest mass are
not assumed to be small. We will write the original
equations from Hwang and Noh (2016) and then
will transform them to a form meeting our goals or,
more specifically, to a form suggesting the spherical
symmetry of the problem and the Lagrangian form.
Quantities like the time $t$, the radial coordinate $r$, and
the matter velocity $v$ are defined in the laboratory
frame, while quantities like the density $\WI{\rho}{b}$, the pressure
$P$, etc. are defined in the comoving frame.
	
\subsection*{Continuity Equation}
	\noindent \noindent The original continuity equation is:
\begin{equation}
\frac{d\rhob}{dt}+\rhob\Big(\mathrm{div}\,\vec{v}+\frac{d\ln\gamma}{dt}\Big)=0.\label{contin_HN}
\end{equation}
Here $\rhob$ is the baryonic density of the matter, $\vec{v}$ is its
velocity, and $\gamma$ is the Lorentz factor:
\begin{equation}
\gamma\equiv\frac{1}{\sqrt{1{-}v^2/c^2}},\label{gamma}
\end{equation}
where $c$ is the speed of light. The total time derivative
is:
\begin{equation}
\frac{d}{dt}=\frac{\partial}{\partial t}+(\vec{v}\cdot\vec{\nabla}).\label{ddt}
\end{equation}
We will rewrite Eq.~(\ref{contin_HN}) as:
\begin{equation}
\frac{d\ln(\rhob\gamma)}{dt}+\frac{1}{r^2}\frac{\partial\left(r^2 v\right)}{\partial r}=0,\label{contin_HN_1}
\end{equation}
where $r$ is the radius (Eulerian coordinate) and the
divergence is written for the case of spherical symmetry.
Let us now introduce a natural definition for the
Lagrangian (mass) coordinate $\mb$ (baryonic mass):
\begin{equation}
\frac{\partial\mb}{\partial r}=4\pi r^2\rhob\gamma,\label{mb}
\end{equation}
where, in comparison with the nonrelativistic case,
the additional factor $\gamma$ appears on the right-hand side.
Let us write this expression as
\begin{equation}
\frac{1}{\rhob\gamma}=\frac{4\pi}{3}\frac{\partial r^3}{\partial\mb}.\label{Continuty}
\end{equation}
Let us show that this is another form of the continuity
equation (\ref{contin_HN_1}). For this purpose, let us differentiate it
with respect to time:
\begin{equation}
-\frac{1}{(\rhob\gamma)^2}\frac{d(\rhob\gamma)}{dt}=4\pi\frac{\partial(r^2 v)}{\partial\mb},
\end{equation}
where we used the fact that $v\equiv dr/dt$. Substituting
here $\partial\mb/\partial r$ from (\ref{mb}), we will obtain exactly (\ref{contin_HN_1}). We
will use (\ref{Continuty}) as the main Lagrangian form of the continuity
equation and $\mb$ as the Lagrangian coordinate.
	
\subsection*{Energy Equation}
\noindent The energy equation in Hwang and Noh (2016) is
written as
\begin{equation}
\frac{d\rho}{dt}+\left(\rho+P/c^2\right)\Big(\mathrm{div}\,\vec{v}+\frac{d\ln\gamma}{dt}\Big)=\WI{R}{E},\label{energy_HN}
\end{equation}
where $P$ is the matter pressure, $\rho=\rhob(1+E/c^2)$ is
the total mass–energy density, and $E$ is the internal
energy of the matter per unit mass. The expression
for the right-hand side $\WI{R}{E}$ is written via the spatial
part $\Pi_{ij}$ of the anisotropic energy–momentum tensor
 $\pi_{\alpha\beta}$, which is related to the total energy–momentum
tensor of the matter $T_{\alpha\beta}$ by the relation
\begin{equation}
T_{\alpha\beta}=(\rho c^2+P)u_\alpha u_\beta+P g_{\alpha\beta}+\pi_{\alpha\beta},
\end{equation}
where $u_\alpha$ is the 4-velocity and $g_{\alpha\beta}$ is a metric tensor
with signature $(-,+,+,+)$. In what follows, we
adopt the condition that the Greek and Latin indices
run the values of $(0{\div}3)$ and $(1{\div}3)$, respectively. We
will write out the specific form of $\WI{R}{E}$ below.

Subtracting the continuity equation (\ref{contin_HN}) from (\ref{energy_HN}),
after simple transformations it is easy to obtain the
energy equation in the form
\begin{equation}
\frac{d E}{dt}+P\frac{d}{dt}\!\left(\!\frac{1}{\rhob}\!\right)=\frac{\WI{R}{E}c^2}{\rhob}.\label{energy_1}
\end{equation}

\subsection*{Equation of Motion}
\noindent The equation of motion is
\begin{equation}
\frac{d\vec{v}}{dt}=\WI{\vec{a}}{G}-\frac{\vec{\nabla} P+\vec{v}\dot{P}/c^2+\WI{\vec{R}}{V}}{\gamma^2(\rho+P/c^2)},\label{motion_eq}
\end{equation}
where $\WI{\vec{a}}{G}=-\vec{\nabla}\varphi$ is the gravitational acceleration
and $\varphi$ is the gravitational potential the equation for
which is given below. The dot in $\dot{P}$ denotes a partial
time derivative and $\WI{\vec{R}}{V}$ is the part of the acceleration
that depends on $\Pi_{ik}$, whose explicit form will be discussed
later.

\section*{ARTIFICIAL VISCOSITY}
\noindent To write the complete system of relativistic hydrodynamics
equations in the final form, it remains to
determine the specific form of the tensor $\Pi_{ij}$. Here we
will consider the contribution to it only from the artificial
viscosity needed to calculate the shock waves.
We will follow the paper by Liebendoerfer et al. (2001),
where the following general relativistic expression for
the viscosity tensor $Q_{\alpha\beta}$ was proposed:
\begin{equation}
Q_{\alpha\beta}=\triangle l^2\rhob u^{\mu}_{;\mu}\Big[\varepsilon_{\alpha\beta}-\frac{1}{3}u^{\mu}_{;\mu}P_{\alpha\beta}\Big],\label{Q_lieben}
\end{equation}
which is valid at $u^{\mu}_{;\mu}<0$, otherwise it is zero. Here,
$\triangle l$ is the characteristic shock front ``smearing'' width,
$P_{\alpha\beta}=u_\alpha u_\beta+g_{\alpha\beta}$, and
\begin{equation}
\varepsilon_{\alpha\beta}=\frac{1}{2}\left(u_{\alpha ;\mu}P^{\mu}_{\beta}+u_{\beta ;\mu}P^{\mu}_{\alpha}\right).\label{epsil_lieben}
\end{equation}
Using the continuity equation (\ref{contin_HN}), the derivative $u^{\mu}_{;\mu}$
can be represented as
\begin{equation}
u^{\mu}_{;\mu}=\frac{1}{c}\frac{\partial\gamma}{\partial t}+\frac{\partial}{\partial x_i}\!\left(\frac{\gamma v_i}{c}\right)=-\frac{\gamma}{c}\frac{d\ln\rhob}{dt}.\label{umumu}
\end{equation}
Hence it can be seen that the artificial viscosity (\ref{Q_lieben})
does ``work'' only during matter compression. To
calculate the remaining quantities, we will use the
spherical symmetry of the problem. We will write
the velocity as $\vec{v}=v \vec{r}/r$ and express the 4-velocity
via $\gamma$ as $u^\alpha=(\gamma,\frac{\vec{r}}{r}\sqrt{\gamma^2{-}1})$.
We will also use the definition (\ref{ddt}) and, as a consequence, the fact that for
any scalar $a$
\begin{equation}
u^\mu a_{;\mu}=\frac{\gamma}{c}\frac{d a}{dt}.
\end{equation}
In addition, we will use the relationship between the
total time derivatives of $\rhob$ and $\gamma$ and the partial
derivative $\partial v/\partial r$ following from (\ref{contin_HN_1}). The final expression
will be written as (recall that for the tensor $\Pi_{ij}$
we need only the spatial components)
\begin{equation}
Q_{ij}=\triangle l^2\rhob\frac{\gamma^2}{3}\frac{d\ln\rhob}{dt}\Big(\frac{d\ln\rhob}{dt}+\frac{3 v}{r}\Big)\Big[(2\gamma^2{+}1)\frac{r_i r_j}{r^2}-\delta_{ij}\Big].
\end{equation}
Since we work in Lagrangian variables, it is convenient
to replace the length scale $\triangle l$ via the characteristic
baryonic mass $\triangle\WI{m}{b}$ using (\ref{mb}). Finally, we will
write
\begin{equation}
\Pi_{ij}=Q_{ij}=\frac{Q}{2}\left[(2\gamma^2{+}1)\frac{r_i r_j}{r^2}-\delta_{ij}\right],\label{PiQ}
\end{equation}
where the factor $1/2$ is introduced for convenience
and $Q$ is the ordinary artificial viscosity, which after
some transformations takes the form
\begin{equation}
Q=-\frac{2}{3}\left(\!\frac{\triangle\WI{m}{b}}{4\pi r^2}\!\right)^2\frac{d}{dt}\!\left(\!\frac{1}{\rhob}\!\right)
 \frac{d}{dt}\Big(\!\ln(\rhob r^3)\!\Big),\label{Qmain}
\end{equation}
for $Q>0$, otherwise $Q=0$. This expression has
the following peculiarity noted in Liebendoerfer et al.
(2001): it becomes zero non only during matter expansion
(the first factor with $d/dt$), but also during
homologous compression (i.e., for $v\propto r$, the second
factor), which allows the nonphysical matter overheating
to be avoided.

\section*{POISSON EQUATION}
\noindent The generalization of the Poisson equation for
the gravitational potential is written as (Hwang and
Noh 2016):
\begin{equation}
\triangle\varphi=4\pi G\Big(\rho+3P/c^2+\frac{2}{c^2}\big[\gamma^2 v^2(\rho+P/c^2)+\Pi^i_i\big]\Big). \label{Puasson}
\end{equation}
Using the explicit form (\ref{PiQ}) of the tensor $\Pi_{ij}$, we can
find its convolution: $\Pi^i_i=(\gamma^2{-}1)Q$. The equation
for the gravitational acceleration $\WI{a}{G}=-\partial\varphi/\partial r$ is
then written as
\begin{equation}
\gamma\frac{\partial\left(r^2\WI{a}{G}\right)}{\partial\mb}={-}G\Big(2\gamma^2{-}1+\frac{(2\gamma^2{-}1)E}{c^2}+\frac{(2\gamma^2{+}1)P}{\rhob c^2}+\frac{2(\gamma^2{-}1)Q}{\rhob c^2}\Big),\label{aG}
\end{equation}
where we took into account the spherical symmetry of
the problem.

\section*{RIGHT-HAND SIDE OF THE ENERGY EQUATION}
\noindent The expression for the right-hand side of the energy
equation $\WI{R}{E}$ is
\begin{equation}
\WI{R}{E}=-\frac{1}{c^2}\Big[\Pi^j_i \nabla_j v^i+\Pi_{ij}\frac{v^i \dot{v}^j}{c^2}\Big],
\end{equation}
where the dot again denotes a partial time derivative.
Writing the velocity as $\vec{v}=v\vec{r}/r$ and using Eq. (\ref{PiQ})
for the tensor $\Pi_{ij}$ , it is easy to obtain
\begin{equation}
\WI{R}{E}=\frac{Q}{c^2}\Big[\frac{v}{r}-\gamma^2\Big(\frac{\partial v}{\partial r}+\frac{v\dot{v}}{c^2}\Big)\Big].
\end{equation}
The expression in square brackets is transformed using
the continuity equation (\ref{contin_HN_1}) into
\begin{equation}
\frac{d\ln\rhob}{dt}+\frac{3v}{r}=\frac{d}{dt}\Big(\!\ln(\rhob r^3)\!\Big).
\end{equation}
The energy balance equation (\ref{energy_1}) will then finally be
written as
\begin{equation}
\frac{dE}{dt}+P\frac{d}{dt}\!\left(\!\frac{1}{\rhob}\!\right)=\frac{Q}{\rhob}\frac{d}{dt}\Big(\!\ln(\rhob r^3)\!\Big).\label{Energy}
\end{equation}
Remarkably, the same factor with the total time
derivative of $\rhob r^3$ as that in the definition of the
artificial viscosity (\ref{Qmain}) appeared on the right-hand
side of (\ref{Energy}).

\section*{RIGHT-HAND SIDE OF THE EQUATION OF MOTION}
\noindent The quantity $\WI{\vec{R}}{V}$ on the right-hand side of the
equation of motion (\ref{motion_eq}) looks as follows\footnote{The erratum in the original paper of Hwang and Noh (2016)
containing the superfluous factor $1/\gamma^2$ at the second term in
square brackets in (\ref{R_V}) was corrected here.}:
\begin{equation}
(\WI{R}{V})_i=\Pi^j_{i,j}-\frac{1}{c^2}\left[v_i(\Pi^k_j v^j)_{,k}-(\Pi_{ij}v^j)^{\cdot}\right]-\frac{v_i}{c^4}(\Pi_{jk}v^j v^k)^{\cdot}.\label{R_V}
\end{equation}
Given the spherical symmetry of the problem and the
explicit form (\ref{PiQ}) for the tensor $\Pi_{ij}$, we will obtain a
number of relations:
\begin{align}
\Pi_{ij}v^j&=\frac{r_i}{r}v\gamma^2 Q,\\
\Pi_{ij}v^jv^j&=v^2\gamma^2 Q=(\gamma^2{-}1)Q c^2.
\end{align}
The first term in (\ref{R_V}) is
\begin{equation}
\Pi_{i,j}^j=\frac{r_i}{r}\Big[\gamma^2\frac{\partial Q}{\partial r}+\Big(\frac{2\gamma^2{+}1}{r}+\frac{\partial\gamma^2}{\partial r}\Big)Q\Big].
\end{equation}
The factor at $v_i$ in the first term in square brackets
(\ref{R_V}) is
\begin{equation}
(\Pi^k_j v^j)_{,k}=\frac{1}{r^2}\frac{\partial}{\partial r}\big(v\gamma^2 r^2 Q\big).
\end{equation}
Collecting all terms, we finally can write
\begin{equation}
(\WI{R}{V})_i=\frac{r_i}{r}\left[\frac{\partial Q}{\partial r}+\frac{3 Q}{r}+\gamma^2\frac{d v}{dt}\frac{Q}{c^2}+\frac{v}{c^2}\frac{\partial Q}{\partial t}\right].
\end{equation}
Substituting this expression into (\ref{motion_eq}), after some
transformations using, in particular, the energy equation
(\ref{Energy}), we will obtain the equation of motion
\begin{equation}
\frac{d}{dt}\Big(\gamma v\big[1+\frac{E{+}(P{+}Q)/\rhob}{c^2}\big]\Big)=\gamma^3\Big(1+\frac{E{+}P/\rhob}{c^2}\Big)\WI{a}{G}-
4\pi r^2\frac{\partial(P{+}Q)}{\partial\mb}-\frac{3Q}{\gamma r\rhob}.\label{motion}
\end{equation}

\section*{DIMENSIONLESS FORM OF THE EQUATIONS}
\noindent For numerical calculations of hydrodynamic processes
in a star it is convenient to have the above
equations in a dimensionless form. We will use the
system of units based on the total mass $\WI{M}{s}$ and initial
radius $\WI{R}{s}$of the star. The units of time, velocity,
density, energy per unit mass, and pressure are
\begin{align}
[t]&=\sqrt{\frac{\WI{R}{s}^3}{G\WI{M}{s}}},\\
[v]&=\sqrt{\frac{G\WI{M}{s}}{\WI{R}{s}}},\\
[\rho]&=\frac{\WI{M}{s}}{4\pi\WI{R}{s}^3},\\
[E]&=\frac{G\WI{M}{s}}{\WI{R}{s}},\\
[P]&=\frac{G\WI{M}{s}^2}{4\pi\WI{R}{s}^4}.
\end{align}
It is also convenient to introduce the relativistic parameter
$\rg$:
\begin{equation}
\rg=\frac{G\WI{M}{s}}{\WI{R}{s}c^2}.
\end{equation}
The parameter $\gamma$ is then
\begin{equation}
\gamma=\frac{1}{\sqrt{1-\rg \hat{v}^2}},\label{gamma_dim}
\end{equation}
where $\hat{v}$ is the dimensionless velocity. The continuity
equation (\ref{Continuty}) in dimensionless variables takes the form
\begin{equation}
\frac{3}{\rhod\gamma}=\frac{\partial \hat{r}^3}{\partial m},\label{Continuty_dim}
\end{equation}
where $\rhod$, $\hat{r}$, and $m$ are the dimensionless density,
radius, and mass coordinate, respectively.

For a further analysis it will be convenient to introduce
the following dimensionless quantities: $\epsilon$ for
the internal energy $E$, $p$ for the ratio $P/\rhob$, and $q$ for
$Q/\rhob$.
The expression for the artificial viscosity (\ref{Qmain}) will
then be written as
\begin{equation}
q=-\frac{(\triangle m)^2}{3\hat{r}^4}\frac{d}{d\tau}\!\left(\!\frac{1}{\rhod^2}\!\right)
 \frac{d}{d\tau}\Big(\!\ln(\rhod \hat{r}^3)\!\Big),\label{Qmain_dim}
\end{equation}
where $\tau$ is the dimensionless time.

The energy equation (\ref{Energy}) will take the form
\begin{equation}
\frac{d\epsilon}{d\tau}-p\frac{d\ln\rhod}{d\tau}=q\frac{d}{d\tau}\Big(\!\ln(\rhod \hat{r}^3)\!\Big).\label{Energy_dim}
\end{equation}

Equation (\ref{aG}) for the gravitational acceleration
will be written as
\begin{equation}
-\gamma\frac{\partial\left(\hat{r}^2\WI{\hat{a}}{G}\right)}{\partial m}=2\gamma^2{-}1+\rg\big[(2\gamma^2{-}1)\epsilon+(2\gamma^2{+}1)p+2(\gamma^2{-}1)q\big].\label{aG_dim}
\end{equation}

And, finally, the equation of motion (\ref{motion}) will take
the form
\begin{equation}
\frac{d}{d\tau}\big(\gamma \hat{v}\big[1+\rg(\epsilon+p+q)\big]\big)=\gamma^3\big(1+\rg(\epsilon+p)\big)\WI{\hat{a}}{G}-
\hat{r}^2\frac{\partial\big((p+q)\rhod\big)}{\partial m}-\frac{3q}{\gamma \hat{r}}.\label{motion_dim}
\end{equation}
Equations (\ref{Continuty_dim}), (\ref{Energy_dim}), and (\ref{motion_dim}) in combination with
the definitions (\ref{gamma_dim}), (\ref{Qmain_dim}), and (\ref{aG_dim}) (and the equality
$\hat{v}=d\hat{r}/d\tau$) comprise the complete system of ideal
relativistic hydrodynamics equations.

\section*{FORMULATION OF THE EXPLOSION PROBLEM}
\noindent In the formulation of the initial conditions in the
problem of the explosion of a minimum-mass neutron
star (MMNS) we will largely follow the paper of Blinnikov
et al. (1990).

For the equation of state (EoS) of the matter we
use a universally accepted, though significantly simplified,
approach: thermodynamic quantities, such as
the pressure and the internal energy, are represented
by sums:
\begin{gather}
P=P_0(\rhob)+\frac{\rhob kT}{\WI{m}{u}}+\frac{a T^4}{3},\label{P_eos}\\
E=E_0(\rhob)+\frac{3kT}{2\WI{m}{u}}+\frac{a T^4}{\rhob},\label{E_eos}
\end{gather}
where $T$ is the temperature, $k$ is the Boltzmann
constant, $\WI{m}{u}$ is the atomic mass unit, and $a=8\pi^5 k^4/15(hc)^3$ is the radiation density constant. The
second and third terms in (\ref{P_eos}--\ref{E_eos}) represent
the contribution of the ideal gas and the radiation,
respectively. For the terms $P_0$ and $E_0$ we use the
fits proposed by Haensel and Potekhin (2004). They
describe the properties of NS matter at temperature
$T=0$ in a wide density range. For our calculations
we chose the BSk22 fit. Naturally, Eqs. (\ref{P_eos}--\ref{E_eos})
describe very roughly the temperature part of the
matter thermodynamics and may be considered only
as the first approximation to reality.

	\begin{figure}[htb]
		\begin{center}
			\includegraphics[width=10cm]{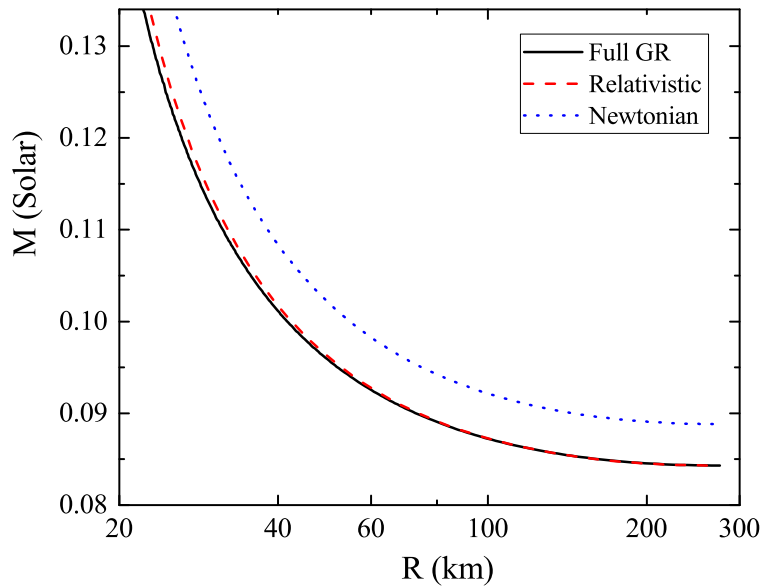}
		\end{center}
		\caption{NS mass–radius diagram near the minimum mass. The blue dotted and black solid lines indicate the nonrelativistic
case and the solution of the GR equations, respectively. The red dashed line indicates our approach.} \label{pix_MR_min}
	\end{figure}
Now it is necessary to determine the MMNS parameters:
for this purpose, we construct a sequence
of NS models parameterized by a decreasing central
density (see, e.g., Haensel et al. 2007) and find the
MMNS in the sequence. The NS matter temperature
is assumed to be zero. Figure~\ref{pix_MR_min} shows the
lower part of the NS mass–radius diagram computed
within several approaches. The blue dotted line indicates
the nonrelativistic case. The black solid line
indicates the solution of the Tolman–Oppenheimer–Volkoff equations (see, e.g., Weinberg 1972). The red
dashed line indicates our approach. As the central
density decreases, the mass of the star drops, while
its radius increases. Each curve in the figure ends
with the MMNS configuration: there are no stable
configurations of lower-mass NSs. As can be seen,
our approach based on the simultaneous solution of
the equilibrium equation (\ref{motion}) and the Poisson equation
(\ref{aG}) at $v=0$ and $Q=0$ is very close to the GR
result, while the Newtonian one gives a slightly higher
MMNS mass at almost the same radius.

\section*{SIMULATION OF EXPLOSIVE MMNS DISRUPTION}
\noindent A MMNS ``explosion'' can be initiated in two
ways: either to remove part of the mass from the
stellar surface or to slightly ``push'' it outward by
specifying an initial small velocity perturbation, for
example, in the form of $v\propto r$. Blinnikov et al. (1990)
showed that both ways are virtually equivalent with
regard to the final result --- the MMNS explosion
parameters. However, we preferred the second variant.
The MMNS envelope is very extended and
rarefied, and by removing even a small fraction of the
mass from the surface, we thus change dramatically
the initial stellar radius. In addition, as will be seen
from our further analysis, the shock breakout and
acceleration on the stellar surface depend strongly on
the density profile in the envelope. The computation of
this process can be distorted by the artificial removal
of part of the envelope or its insufficient numerical
resolution in the computation.

For a better presentation of our numerical simulation
results, we divided the illustration of the MMNS
explosion process into three periods. The first period
lasts from the loss of stability by the star to the formation
of a shock front in the outer part of the envelope.
Next, in the second period, we show in detail the
shock breakout whereby part of the ejected matter is
accelerated significantly. The last, third period shows
the expansion of matter and the establishment of a
final ejecta velocity distribution.

\begin{figure}[htb]
		\begin{center}
			\includegraphics[width=\linewidth]{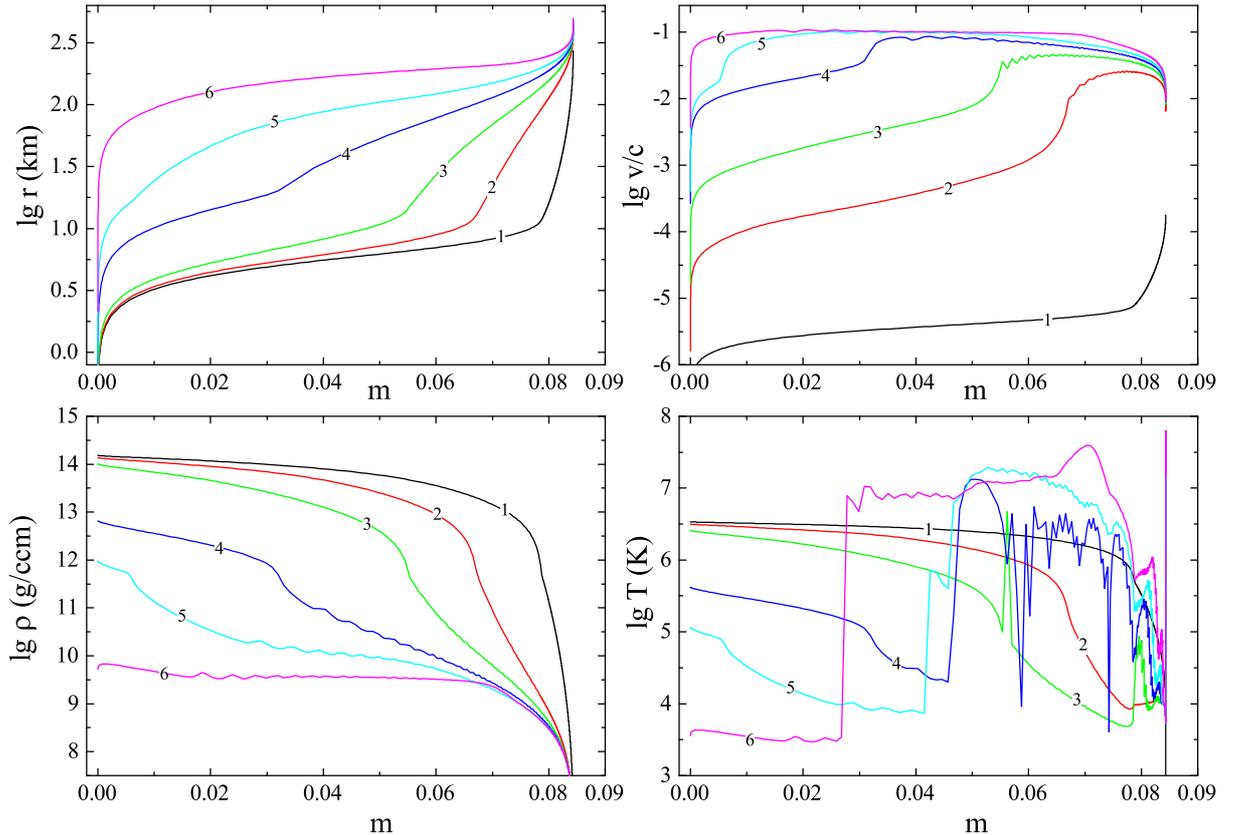}
		\end{center}
		\caption{The first MMNS explosion period: from the equilibrium configuration (index 1) to the time of shock generation (index 6).
The logarithms of the Eulerian coordinate $r$ (top left), the ratio $v/c$ (top right), the density (bottom left), and the temperature
(bottom right) are shown as a function of the mass coordinate $m$.} \label{pix_expl_before}
	\end{figure}
The development of the explosion in the first period
is shown in Fig.~\ref{pix_expl_before}. It presents the distributions of
the radius $r$, the ratio $v/c$, the density $\rho$, and the
temperature $T$ as functions of the mass coordinate $m[M_\odot]$
 for six times (the indices near the curves).
On the graphs from the second to the fourth one the
times corresponding to the indices are given in Table
1. The initial small velocity perturbation (time 1)
corresponds to $v\propto r$. As can be seen, the evolution
over times 1–6 leads to a flattening of the density and
velocity distributions. Hydrodynamic disturbances
(waves) are generated in this case, which are clearly
seen, for example, on the velocity graphs at times 3
and 4 and in the density distributions 4–6. Propagating
outward along a descending density profile,
these disturbances transform into weak shock waves
and lead to envelope heating, which is clearly seen
on the temperature panel. By time 6 at a mass
coordinate $m\approx 0.07$ evidence for the generation of
a strong shock wave can be seen on the velocity
and, especially, temperature graphs, which we will
consider below.

\begin{table}[t]
\vspace{6mm}
\centering
{{\bf Table 1.} The times for the indices from 1 to 12 in Figs \ref{pix_expl_before}--\ref{pix_expl_after}}
\vspace{5mm}
\begin{tabular}{|c|c|c|c|c|c|c|} \hline\hline
Index &1 & 2 & 3 & 4 & 5 & 6\\
\hline
Time (s) & 0.0 & 0.3110 & 0.3161 & 0.3191 & 0.3210 & 0.3236\\
\hline\hline
Index & 7 & 8 & 9 & 10 & 11 & 12\\
\hline
Time (s)  & 0.3252 & 0.3267 & 0.3292 & 0.3307 & 0.3573 & 0.6214\\
\hline\hline
\end{tabular}
\end{table}

\begin{figure}[htb]
		\begin{center}
			\includegraphics[width=\linewidth]{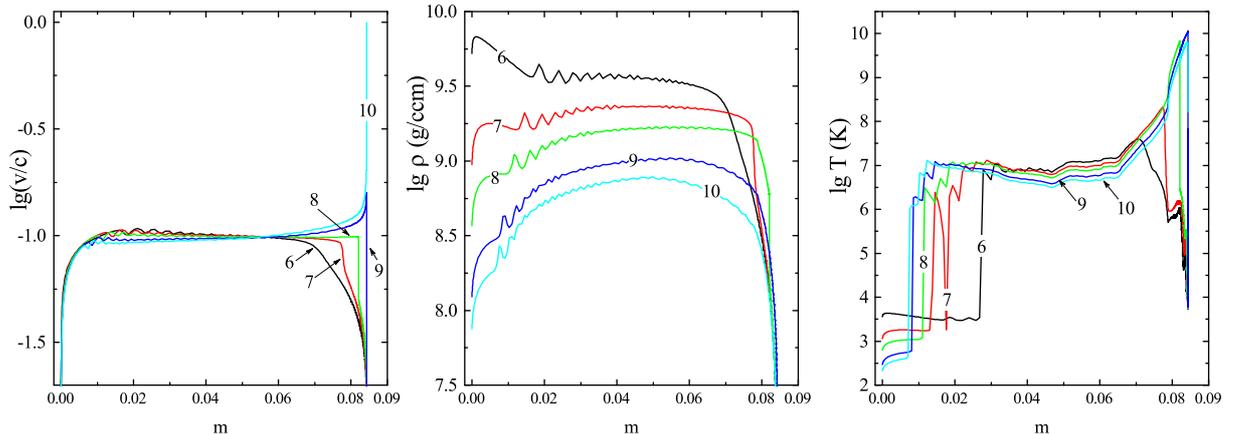}
		\end{center}
		\caption{The second MMNS explosion period: shock breakout. The velocity $v$, density $\rho$, and temperature $T$ of the matter are
shown as functions of the mass coordinate $m$.} \label{pix_expl_swb}
	\end{figure}
The shock generation and breakout are shown
in Fig.~\ref{pix_expl_swb}. The shock front and its propagation are
clearly seen on the velocity panel (left). The shock
passage through the envelope causes its temperature
to rise by more than two orders of magnitude (right),
reaching $T\sim 10^{10}$~K (i.e., $\sim 1$MeV). Furthermore,
at shock breakout the outermost layers of the envelope
are accelerated to ultrarelativistic velocities
due to the cumulation effect (time 10 on the velocity
graph). However, the fraction of the mass accelerated
to $v/c\sim 1$ is extremely small (see Fig.~\ref{pix_VM} below and its
discussion).

\begin{figure}[htb]
		\begin{center}
			\includegraphics[width=\linewidth]{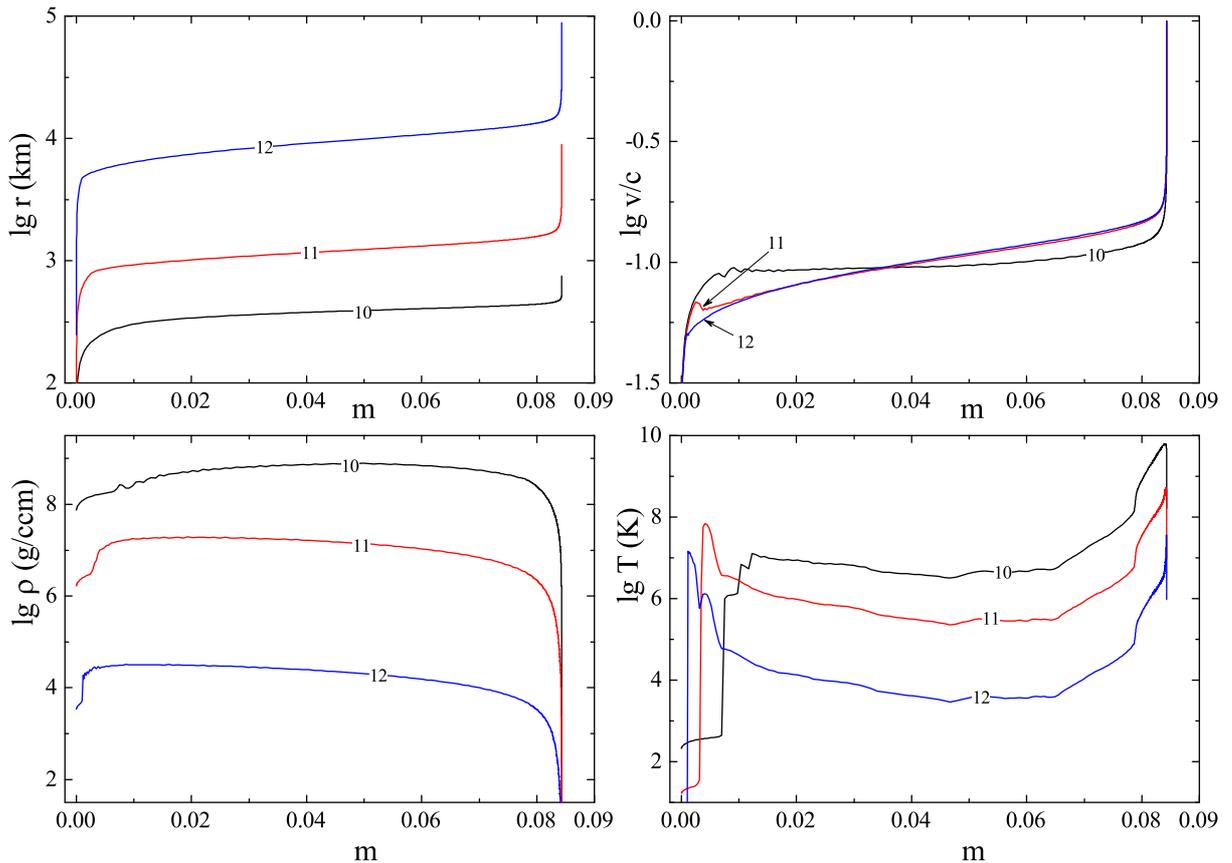}
		\end{center}
		\caption{The third MMNS explosion period: from shock breakout to free expansion. The notation is the same as that in Fig.~\ref{pix_expl_before}.} \label{pix_expl_after}
	\end{figure}
The third explosion period from shock breakout to
the transition to free expansion is illustrated in Fig.~\ref{pix_expl_after}.
As can be seen, the star expands essentially at a
density uniform in $m$. The velocity comes fairly rapidly
to its final distribution (time 12). The temperature
decreases monotonically virtually in the entire volume
of the star, except for its inner part, where weak
shocks lead to some additional heating.

\section*{DISCUSSION AND CONCLUSIONS}
\noindent It is useful to compare the MMNS explosion parameters
given above both with the original computation
from Blinnikov et al. (1990) and with the
computation in which an identical problem is solved
within ordinary Newtonian hydrodynamics. Table~2
contains a comparison, where Blinn denotes the results
from the mentioned 1990 paper, non-rel is the
nonrelativistic case, and rel is the relativistic one.
\begin{table}[t]
\vspace{6mm}
\centering
{{\bf Table 2.} Comparison of the computational results for several
approaches.}

\vspace{5mm}\begin{tabular}{|c|c|c|c|} \hline\hline
 & Blinn & non-rel & rel\\
 \hline
$\WI{M}{min},~M_\odot$ & 0.095 & 0.089 & 0.084\\
\hline
$\WI{N}{zone}$ & 151 & 2609 & 3899\\
\hline
$\WI{E}{exp},~10^{50}$~эрг & 8.8 & 9.1 & 8.7\\
\hline\hline
\end{tabular}
\end{table}

The first row $\WI{M}{min}$ gives the mass of the exploding
NS. The difference between our Newtonian computation
and the computation by Blinnikov et al. (1990)
is due to the use of different approximations for $P_0(\rho)$
and $E_0(\rho)$ (see Eqs. \ref{P_eos} and \ref{E_eos}). The difference
between non-rel and rel consists in using the Newtonian/
relativistic stellar equilibrium equations.

The number of zones $\WI{N}{zone}$ in the Blinn computation
is more than an order of magnitude smaller than
that in our one. This did not allow one to provide a
good resolution of the extended NS envelope and to
correctly compute the cumulation of the shock during
its breakout. That is why in this computation we did
not run into the problem of $v/c>1$ discussed below.

Finally, the last row in the table gives the kinetic
energy $\WI{E}{kin}$ of the ejecta at the end of the computation.
It can be seen that all three values are very
close, while the differences between rel and non-rel
are due mainly to the difference in the total masses
(see also Fig.~\ref{pix_VM} below). On the whole, however,
all three computations show a very similar picture of
explosion development.

\begin{figure}[htb]
		\begin{center}
			\includegraphics[width=\linewidth]{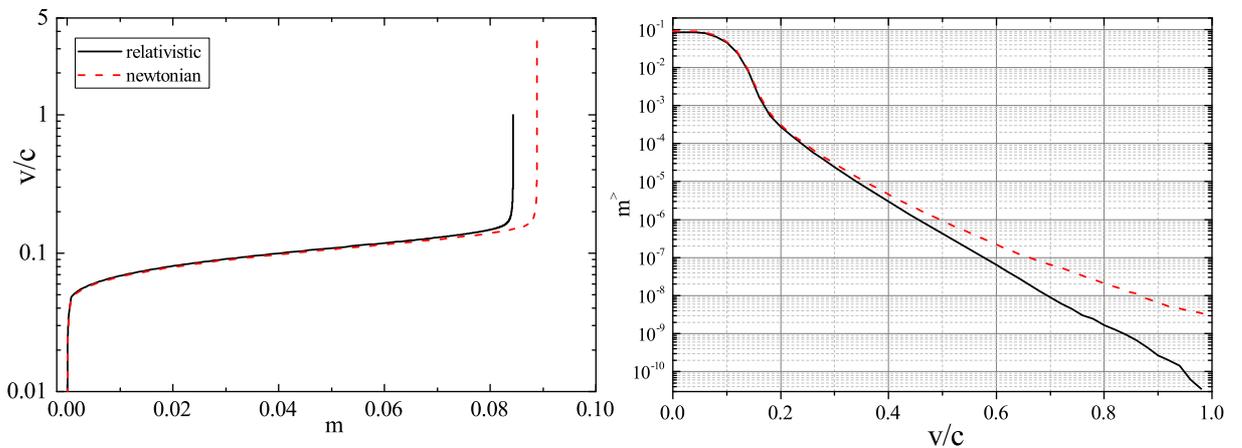}
		\end{center}
		\caption{Comparison of the relativistic and Newtonian calculations. Left: The velocity distribution as a function of m at the final
time. Right: The $m^{>}(v/c)$ distribution (for more details, see the text).} \label{pix_VM}
	\end{figure}
Let us now turn to the problem that gave rise
to this paper. Figure~\ref{pix_VM} (left) shows the ejecta velocity
distribution at the final time of our computations.
As can be seen, in the nonrelativistic case part of
the matter (though very small) was accelerated to
$v/c>1$. Otherwise, however, the distributions are
surprisingly similar.

Figure\ref{pix_VM} (right) shows the $m^{>}$ distribution as a function
of the parameter $v/c$. By definition, $m^{>}(v/c)$ is the
ejecta mass (in units of the solar mass) that has
a velocity greater than $v/c$ at the final time of our
computation. For example, the mass accelerated to
$v/c>0.1$ is approximately $m^{>}\approx 0.047$ (i.e., almost
half the entire stellar mass), while the mass accelerated
to $v/c>0.2$ is already only $3\times 10^{-4}$. It can be
seen from the same graph that the mass accelerated
to $v/c>1$ in the nonrelativistic computation is $\sim 3\times 10^{-9}$. Note that the ejecta velocity distribution
can be important for a comparison with the results
of observations, in particular, for determining the
properties of the so-called red and blue kilonovae for
GRB170817A (Siegel 2019).

In conclusion, note that, on the whole, the MMNS
explosion parameters derived by D.K.~Nadyozhin
in 1990 (Blinnikov et al. 1990) within the nonrelativistic
approach agree well both with our results in
the same approximation and with the completely relativistic
computation (see, e.g., Table 2 and left part of Fig.~\ref{pix_VM}).
At the same time, it should be emphasized that in our
computations we omitted several important points
whose inclusion can affect the results obtained. First,
the energy losses through neutrino radiation, for
example, from electron–positron pair annihilation,
become important in the region heated by the shock
front with a temperature $T\sim 10^{10}$~K. These losses
can reduce the cumulation effect and, accordingly,
the ejecta velocity. This phenomenon is planned to be
studied in the near future.

Second, during the disruption of a MMNS its
matter experiences explosive decompression, which
will be accompanied by numerous neutron-capture
and beta-decay processes (r-process) leading to a
significant change in the nuclear composition of the
matter. At the same time, neutrinos are also emitted,
but the energy of the corresponding nuclear transformations
is released as well. Our preliminary offline
computations (Panov and Yudin 2020) on fixed
tracks show the promise and potential importance of
these processes, which will undoubtedly also need to
be studied in detail in future.

Third, it should be remembered that our formulation
of the problem is idealized. In reality, a low-mass
NS is a member of a binary system. As a result of
mass transfer, it not only loses its mass, but is also
heated due to the tidal interaction with its companion.
In addition, as our preliminary computations show,
the stability of the transfer process is lost not at
$M_2=\WI{M}{min}\sim 0.1 M_\odot$, but at $M_2\sim 0.2 M_\odot$. The star
loses the rest of its mass on a fast hydrodynamic time
scale, and the initial configuration for the MMNS
explosion can differ noticeably from the spherically
symmetric one already due to the influence of gravity
from its massive companion (Manukovskii 2010).
Computing the MMNS explosion in this formulation
is a challenging three-dimensional problem that is of
importance in its own right.

\vspace{1cm}
 {\bf ACKNOWLEDGMENTS}

This work was supported by the Russian Foundation
for Basic Research (project no. 18-29-21019mk). Author is grateful to the anonymous referees whose remarks
contributed significantly to an improvement of this paper.

\appendix
\section*{APPENDIX: ENERGY CONSERVATION IN A STAR}
\noindent The relativistic equations derived by us look unusual
with regard to how the terms with the artificial
viscosity $Q$ appear in them. Usually, the introduction
of the latter is actually reduced to an additive to the
pressure: $P\rightarrow P{+}Q$. In our case, however, there
is also an additional contribution to the equation of
motion (the last turn in (\ref{motion})). Furthermore, the righthand
side of the energy equation (\ref{Energy}) is not reduced to
the form $Q/\rhob^2 (d\rhob/dt)$ because of the term $r^3$ under
the logarithm. Let us demonstrate that, nevertheless,
our formulas are quite self-consistent. We will work
in the nonrelativistic limit. Let us multiply Eq. (\ref{motion})
by $v$ and integrate over $\mb$ throughout the star. On the
right we will obtain the term
\begin{equation}
\frac{1}{2}\int\frac{d v^2}{dt}d\mb=\frac{d\WI{E}{kin}}{dt},
\end{equation}
i.e., the total time derivative of the kinetic energy.
The gravitational acceleration (\ref{aG}) in the limit under
consideration is simply $\WI{a}{G}=-G\mb/r^2$. Its integral is
\begin{equation}
-\int\!\frac{v G\mb d\mb}{r^2}=\int\!\frac{d}{dt}\!\left(\frac{1}{r}\right)G\mb d\mb=-\frac{d\WI{E}{grav}}{dt},
\end{equation}
where $\WI{E}{grav}$ is the gravitational energy. We integrate
the next term in (\ref{motion}) by parts:
\begin{equation}
-\int\! 4\pi r^2 v\frac{\partial (P{+}Q)}{\partial\mb}d\mb=\int\!(P{+}Q)\frac{d}{dt}\!\left(\!\frac{1}{\rhob}\!\right)d\mb.\label{app_intP}
\end{equation}
The energy equation (\ref{Energy}) can be rewritten in an
equivalent form:
\begin{equation}
(P+Q)\frac{d}{dt}\!\left(\!\frac{1}{\rhob}\!\right)=-\frac{dE}{dt}+\frac{3Qv}{r\rhob}.\label{app_PQ}
\end{equation}
Substituting this expression into (\ref{app_intP}) , we will find that
the second term on the right in (\ref{app_PQ}) is canceled out
with the last term in the equation of motion (\ref{motion})
during its integration. There remains only the total
derivative of the star’s internal energy $\WI{E}{int}=\int\! E d\mb$.
Thus,we obtained the law of conservation of the star’s
total energy in the form $d\WI{E}{tot}/dt=0$, where$\WI{E}{tot}=\WI{E}{kin}+\WI{E}{grav}+\WI{E}{int}$.

	\pagebreak

\end{document}